\documentclass{ws-procs9x6}

\begin{document}

\title{Monte Carlo simulations of radio emission from cosmic ray air showers}

\author{T. HUEGE}

\address{Institut f\"ur Kernphysik, \\
Forschungszentrum Karlsruhe, \\ 
Postfach 3640, \\
76021 Karlsruhe, Germany\\ 
E-mail: tim.huege@ik.fzk.de}

\author{H. FALCKE}

\address{ASTRON, \\ 
P.O. Box 2, \\
7990 AA Dwingeloo, The Netherlands\\
E-mail: falcke@astron.nl}  

\maketitle

\abstracts{
As a basis for the interpretation of data gathered by LOPES and other experiments, we have carried out Monte Carlo simulations of geosynchrotron radio emission from cosmic ray air showers. The simulations, having been verified carefully with analytical calculations, reveal a wealth of information on the characteristics of the radio signal and their dependence on specific air shower parameters. In this article, we review the spatial characteristics of the radio emission, its predicted frequency spectrum and its dependence on important air shower parameters such as the shower zenith angle, the primary particle energy and the depth of the shower maximum, which can in turn be related to the nature of the primary particle.}

\section{The simulations}

Two main mechanisms have been considered to contribute to radio emission from cosmic ray air showers: Askaryan-type \v{C}erenkov radiation arising from a negative charge excess moving through the atmosphere at velocities faster than the speed of light in air, and emission generated as a consequence of the deflection of charged particles in the earth's magnetic field. A number of historical results illustrated that, while the \v{Cerenkov} emission mechanism dominates in dense media such as ice, the geomagnetic mechanism is dominant in the atmosphere. Our simulations thus focus on the latter mechanism, interpreting the radio emission as ``coherent geosynchrotron radiation'' arising from the geomagnetic deflection of highly relativistic electron-positron pairs generated in the air shower cascade\cite{FalckeGorham}.

In order to gain a solid understanding of the emission characteristics, in a first step, we have performed analytical calculations of the expected radio signal\cite{Analytics}. In a second step, we have then improved on the analytical results with detailed Monte Carlo simulations, which we have directly compared and thus verified with the analytical results and the available historical data\cite{MonteCarlo}. While these Monte Carlo simulations still incorporate a somewhat simplified air shower model based on analytical parametrizations, they do already take into account the most important air shower characteristics such as longitudinal (arrival time) and lateral particle distributions, energy and track-length distributions, the overall longitudinal development of the air shower and the geometry of the air shower and magnetic field. Our simulations thus for the first time have provided a prediction of the radio emission from realistically modeled air showers. Furthermore, the model is currently being enhanced by substituting the parametrized particle distributions with CORSIKA\cite{CORSIKA}-generated distributions.

We here present only a subset of the results derived with our Monte Carlo code. A more detailed analysis, including a parametrization of the derived dependences, has been published elsewhere\cite{Results}.

\section{Simulation results}

We first consider the very simple scenario of a vertical air shower with primary particle energy of $10^{17}\,$eV developing to its maximum at $\sim630\,$g$\,$cm$^{-2}$, i.e., at $\sim4000\,$m above ground. The geomagnetic field is adopted as 70$^{\circ}$ inclined with a strength of 0.5$\,$G, corresponding approximately to the field configuration in central Europe.

\begin{figure}[ht]
\centerline{\epsfxsize=4.3in\epsfbox{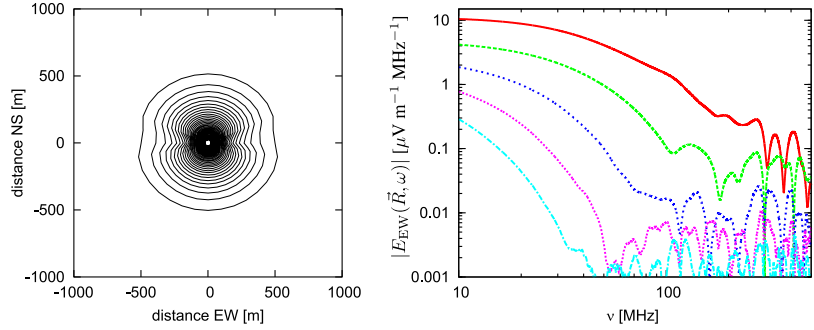}}   
\caption{Radio emission from a vertical $10^{17}\,$eV air shower. Left: $10\,$MHz total field strength emission pattern. Right: Frequency spectra at (from top to bottom) 20$\,$m, 140$\,$m, 260$\,$m, 380$\,$m and 500$\,$m north of the shower center.\label{vertical}}
\end{figure}

Figure \ref{vertical} shows the ground-level total field strength emission pattern at $10\,$MHz, visualized as a contour plot, and the frequency spectra derived at various radial distances from the shower center. The emission pattern shows remarkable symmetry and is almost circular. This is not a trivial result, as the emission process itself, i.e., the deflection of electrons and positrons in the geomagnetic field, is a highly directed process. The circularity of the footprint illustrates that most of the emission stems from particles having short track lengths. A slight north-south asymmetry introduced by the inclination of the geomagnetic field is also visible. The frequency spectra shown in the right panel illustrate that the field strength drops quickly to higher observing frequencies. This is a direct consequence of diminishing coherence as the wavelength becomes shorter and thus comparable to the dimensions of the air shower pancake, in particular its thickness of a few meters. The decrease is stronger at larger distances from the shower center. When one enters the incoherent regime the frequency spectra exhibit unphysical seeming features such as rapidly alternating series of maxima and minima. Realistic calculations of the emission in this regime can only be performed with a better underlying air shower model taking into account inhomogeneities in the shower in very mildly thinned calculations.

\begin{figure}[!b]
\centerline{\epsfxsize=4.3in\epsfbox{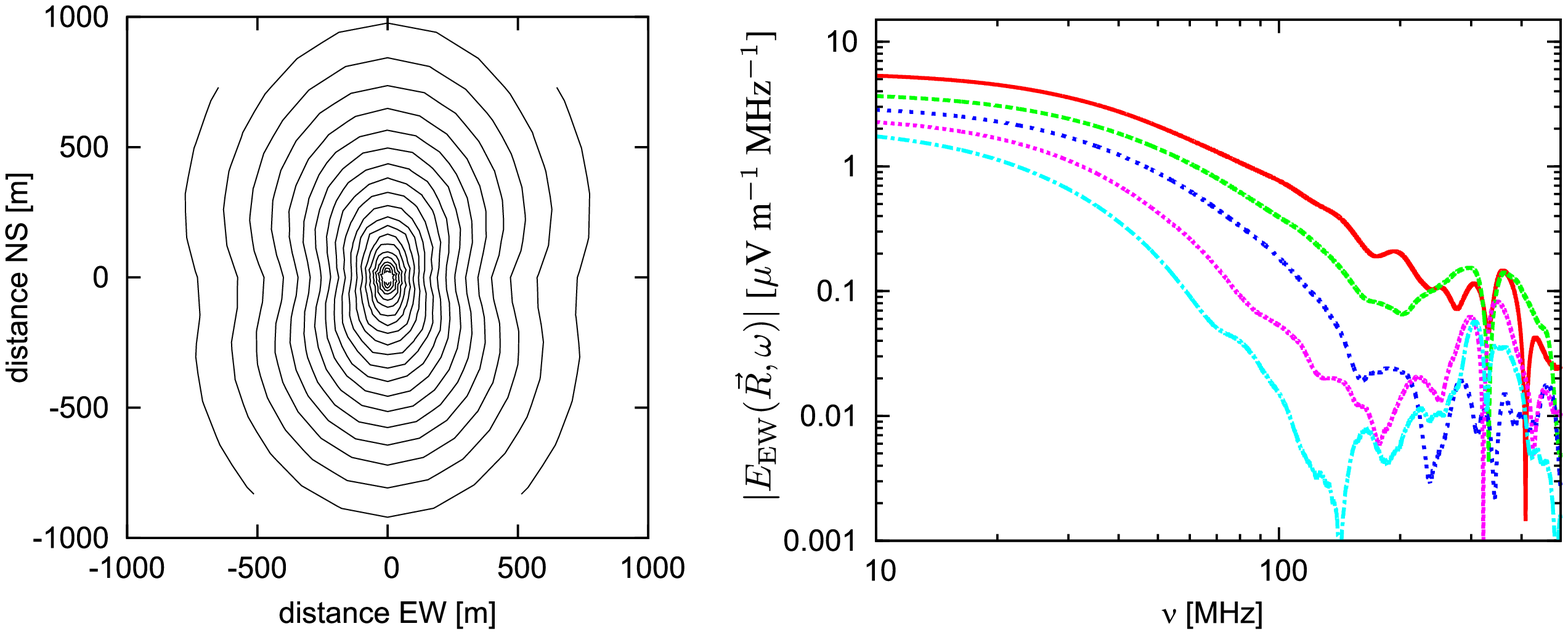}}   
\caption{Radio emission from a 45$^{\circ}$ inclined $10^{17}\,$eV air shower. Left: $10\,$MHz total field strength emission pattern. Right: Frequency spectra at (from top to bottom) 20$\,$m, 140$\,$m, 260$\,$m, 380$\,$m and 500$\,$m north of the shower center.\label{inclined}}
\end{figure}

Figure \ref{inclined} demonstrates the changes arising in the transition from a vertical to a 45$^{\circ}$ inclined air shower (coming from the south). The emission pattern becomes elongated considerably along the shower axis. This is mainly a projection effect directly associated with the inclination of the shower axis. On closer look, however, the emission pattern becomes wider (and less peaked) as a whole, even in the direction perpendicular to the shower axis. The reason for this is that the maximum of the inclined shower located at the same (slant) atmospheric depth of $\sim630\,$g$\,$cm$^{-2}$ now is at much greater geometrical distance from the observer at ground-level. This geometric effect has direct influence on the slope of the radio emission's lateral distribution. A look at the frequency spectra in the right panel shows that coherence is also retained up to higher frequencies in case of inclined showers. Their larger radio footprint combined with the large solid angle associated with medium to high zenith angles thus makes inclined showers a particularly interesting target for radio observations\cite{Results,Franzosen}. At near horizontal inclination, even neutrino-induced air showers might become observable.

\begin{figure}[!b]
\centerline{\epsfxsize=4.5in\epsfbox{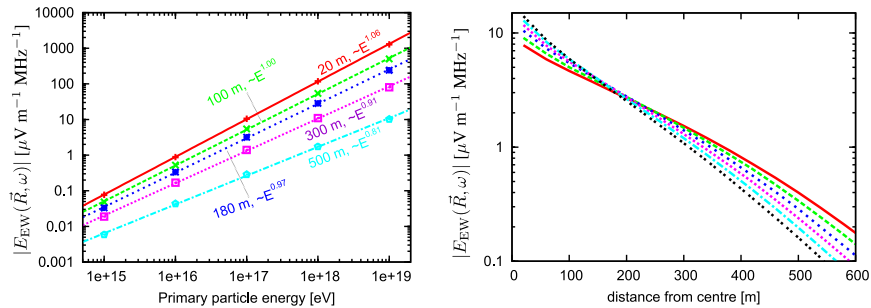}}   
\caption{Radio emission from vertical air showers. Left: Scaling of the $10\,$MHz electric field strength with primary particle energy at various distances from the shower center. Right: Dependence of the radio signal's lateral distribution on the depth of shower maximum $X_{\mathrm{max}}$. $X_{\mathrm{max}}$ in g~cm$^{-2}$ is for red/solid: 560, green/dashed: 595, blue/dotted: 631, violet/short dotted: 665, turquois/dash-dotted: 700, black/double-dotted: 735.\label{energyandxmax}}
\end{figure}

Figure \ref{energyandxmax} illustrates two additional parameters that have direct influence on the radio signal. The left panel shows the impact of the primary particle's energy. The electric field strength at all distances scales as a power-law with the primary particle energy. The power-law index is very close to unity, i.e., that of the linear relation expected for coherent emission. To larger distances, the slope of the power-law gets flatter due to the effect that more energetic showers on average penetrate deeper into the atmosphere and thus have their shower maximum geometrically closer to the observer. As already discussed in the context of inclined showers and illustrated in the right panel, this directly influences the lateral distribution of the radio emission. Since the depth of the shower maximum can in turn be related to the nature of the primary particle, its influence on the radio signal's lateral distribution can potentially be used to probe the primary particle composition with radio measurements\cite{ICRC}.

Another important result of the simulations (not shown here explicitly) are the predicted linear polarization characteristics of the radio signal\cite{Results}. They can be used to directly verify the geomagnetic origin of the emission. To make the simulation results available for easy comparison with experimental data, they are also available as a parametrization formula\cite{Results}.

\section{Conclusions}

We have carried out elaborate Monte Carlo simulations of geosynchrotron radio emission from cosmic ray air showers. Special care has been taken to verify the Monte Carlo results with analytical calculations and historical data, giving us a good understanding of the emission process and thus solid confidence in the predictions. The simulations predict many important characteristics of the radio emission and their relation to parameters of the associated air shower. The total field strength emission pattern is very regular and symmetric in the coherent regime. The geomagnetic origin of the emission can be directly verified with polarization measurements. The frequency spectra cut off quickly to high frequencies, making low observing frequencies around a few tens of MHz desirable. Inclined air showers have a much wider emission pattern and are thus particularly suitable for radio observations. The slope of the lateral distribution can be directly related to the geometrical distance between observer and shower maximum. It is thus not only sensitive to the shower zenith angle, but also to the nature of the primary particle. As expected for coherent emission, the electric field strength scales approximately linearly with the primary particle energy. These predictions will allow to analyze and interpret experimental data such as those provided by LOPES\cite{LOPES} and other experiments.


\end{document}